\documentclass[a4paper,oneside,onecolumn]{article}%
\usepackage{amssymb}
\usepackage{amsmath}
\usepackage{cite}
\usepackage{amscd}
\usepackage{amsfonts}%
\providecommand{\U}[1]{\protect\rule{.1in}{.1in}}
\begin{document}

\title{Laguerre-Gaussian modes: entangled state representation and generalized Wigner
transform in quantum optics}
\author{{\small Li-yun Hu}$^{1}${\small \thanks{Corresponding author. \emph{E-mail
addresses}: hlyun2008@126.com, hlyun2008@gmail.com)} and Hong-yi Fan}$^{2}$\\$^{1}${\small College of Physics \& Communication Electronics, Jiangxi Normal
University, Nanchang 330022, China}\\$^{2}${\small Department of Physics, Shanghai Jiao Tong University, Shanghai
200030, China}}
\date{}
\maketitle

\begin{abstract}
{\small By introducing a new entangled state representation, we show that the
Laguerre-Gaussian (LG) mode is just the wave function of the common
eigenvector of the orbital angular momentum and the total photon number
operators of 2-d oscillator, which can be generated by 50:50 beam splitter
with the phase difference }$\phi=\pi/2$ {\small between the reflected and
transmitted fields. Based on this and using the Weyl ordering invariance under
similar transforms, the Wigner representation of LG is directly obtained,
which can be considered as the generalized Wigner transform of Hermite
Gaussian modes.}



\end{abstract}

\section{Introduction}

It has been known that a Laguerre-Gaussian (LG) beam of paraxial light has a
well-defined orbital angular momentum \cite{1,2,3,4,5}, which is useful in
studying quantum entanglement \cite{6}. In Ref. \cite{3} Nienhuis and Allen
employed operator algebra to describe the Laguerre-Gaussian beam, and noticed
that Laguerre-Gaussian modes are laser mode analog of the angular momentum
eigenstates of the isotropic 2-d harmonic oscillator. In Ref. \cite{7} Simon
and Agarwal presented a phase-space description (the Wigner function) of the
LG mode by exploiting the underlying phase-space symmetry. In this Letter we
shall go a step further to show that LG mode is just the wave function of the
common eigenvector $\left\vert n,l\right\rangle $ of the orbital angular
momentum operator and the total photon number operator of 2-d oscillator in
the{\small \ }entangled state representation (ESR). The ESR was constructed
\cite{8,9} based on the Einstein-Podolsky-Rosen quantum entanglement
\cite{10}. It is shown that $\left\vert n,l\right\rangle $ can be generated by
50:50 beam splitter with the phase difference $\phi=\pi/2$ between the
reflected and transmitted fields. Then we use the Weyl ordering form of the
Wigner operator and the Weyl ordering's covariance under similar
transformations to directly derive the Wigner representation of LG beams,
which seems economical. The marginal distributions of Wigner function (WF) are
also obtained by the entangled state representation. It is found that the
amplitude of marginal distribution is just the eigenfunction of the fractional
Fourier transform (FrFT). In addition, LG mode can also be considered as the
generalized Wigner transform of Hermite Gaussian modes by using the Schmidt
decomposition of the ESR.

\section{Eigenvector corresponding to Laguerre-Guassian mode}

In Ref. \cite{3} the Bosonic operator algebra of the quantum harmonic
oscillator is applied to the description of Gaussian modes of a laser beam,
i.e., a paraxial beam of light is described by operators' eigenvector
equations%
\begin{align}
N\left\vert n,l\right\rangle  &  =n\left\vert n,l\right\rangle ,\text{\ \ }%
N\equiv\left(  a_{1}^{\dagger}a_{1}+a_{2}^{\dagger}a_{2}\right)  ,\nonumber\\
L\left\vert n,l\right\rangle  &  =l\left\vert n,l\right\rangle ,\text{
\ }L\equiv X_{1}P_{2}-X_{2}P_{1}, \label{7}%
\end{align}
since $\left[  N,L\right]  =0$, where $a_{i}^{\dagger}$ and $a_{i}$ ($i=1,2$)
are Bose creation operator and annihilation operator; $L$ and $N$ are the
orbital angular momentum operator and the total photon number operator of a
paraxial beam of light, respectively. Using $X_{i}=\left(  a_{i}%
+a_{i}^{\dagger}\right)  /\sqrt{2}$ and $P_{i}=\left(  a_{i}-a_{i}^{\dagger
}\right)  /(\mathtt{i}\sqrt{2})$, and $[a_{i},a_{j}^{\dagger}]=\delta_{ij}$,
then
\begin{equation}
L=\mathtt{i}(a_{2}^{\dagger}a_{1}-a_{1}^{\dagger}a_{2}). \label{10}%
\end{equation}
Here we search for the common eigenvector of $\left(  N,L\right)  $ in the
entangled state representation. By introducing
\begin{equation}
A_{+}=\frac{1}{\sqrt{2}}(a_{1}-ia_{2}),\text{ }A_{-}=\frac{1}{\sqrt{2}i}%
(a_{1}+ia_{2}), \label{12}%
\end{equation}
which obey the commutative relation
\begin{align}
\lbrack A_{+},A_{+}^{\dagger}]  &  =1,\text{\ }[A_{-},A_{-}^{\dagger
}]=1,\label{13}\\
\lbrack A_{+},A_{-}^{\dagger}]  &  =0,\text{ }[A_{-},A_{+}^{\dagger
}]=0,\nonumber
\end{align}
one can see
\begin{equation}
N=A_{+}^{\dagger}A_{+}+A_{-}^{\dagger}A_{-},\text{ }L=A_{+}^{\dagger}%
A_{+}-A_{-}^{\dagger}A_{-}. \label{14}%
\end{equation}
Now we introduce the entangled state representation in Fock space,
\begin{equation}
\left\vert \eta\right\rangle =\exp\left\{  -\frac{1}{2}\left\vert
\eta\right\vert ^{2}+\eta A_{+}^{\dagger}-\eta^{\ast}A_{-}^{\dagger}%
+A_{+}^{\dagger}A_{-}^{\dagger}\right\}  \left\vert 00\right\rangle ,
\label{15}%
\end{equation}
here $\eta=\left\vert \eta\right\vert e^{i\varphi}=\eta_{1}+i\eta_{2},$
$\left\vert 00\right\rangle $ is annihilated by $A_{+}$ and $A_{-}.$ It is not
difficult to see that $\left\vert \eta\right\rangle $ is the common
eigenvector of operators $\left(  X_{1}-X_{2}-P_{1}+P_{2},P_{1}+P_{2}%
-X_{1}-X_{2}\right)  $
\begin{align}
\left(  X_{1}-X_{2}-P_{1}+P_{2}\right)  \left\vert \eta\right\rangle  &
=2\eta_{1}\left\vert \eta\right\rangle ,\nonumber\\
\left(  P_{1}+P_{2}-X_{1}-X_{2}\right)  \left\vert \eta\right\rangle  &
=2\eta_{2}\left\vert \eta\right\rangle . \label{15a}%
\end{align}
Using the normal ordering form of vacuum projector
\begin{equation}
\left\vert 00\right\rangle \left\langle 00\right\vert =\colon\exp\left(
-A_{+}^{\dagger}A_{+}-A_{-}^{\dagger}A_{-}\right)  \colon, \label{14a}%
\end{equation}
(where : : denotes normal ordering) and the technique of integral within an
ordered product (IWOP) of operators \cite{11,12} we can prove the completeness
relation and the orthonormal property of $\left\vert \eta\right\rangle ,$%
\begin{equation}
\int\frac{d^{2}\eta}{\pi}\left\vert \eta\right\rangle \left\langle
\eta\right\vert =1,\text{ }\left\langle \eta\right.  \left\vert \eta^{\prime
}\right\rangle =\pi\delta\left(  \eta-\eta^{\prime\ast}\right)  \delta\left(
\eta^{\ast}-\eta^{\prime}\right)  . \label{16}%
\end{equation}
\ Thus $\left\vert \eta\right\rangle $ is qualified to make up a new
representation. It follows from (\ref{15}) and (\ref{13}) that
\begin{equation}
A_{+}\left\vert \eta\right\rangle =(\eta+A_{-}^{\dagger})\left\vert
\eta\right\rangle ,\text{\ }A_{+}^{\dagger}\left\vert \eta\right\rangle
=\left(  \frac{\partial}{\partial\eta}+\frac{\eta^{\ast}}{2}\right)
\left\vert \eta\right\rangle , \label{16b}%
\end{equation}%
\begin{equation}
A_{-}\left\vert \eta\right\rangle =(A_{+}^{\dagger}-\eta^{\ast})\left\vert
\eta\right\rangle ,A_{-}^{\dagger}\left\vert \eta\right\rangle =\left(
-\frac{\partial}{\partial\eta^{\ast}}-\frac{\eta}{2}\right)  \left\vert
\eta\right\rangle . \label{17}%
\end{equation}
which lead to (denote $r=\left\vert \eta\right\vert $ for simplicity)
\begin{align}
(A_{+}^{\dagger}A_{+}+A_{-}^{\dagger}A_{-})\left\vert \eta\right\rangle  &
=\left(  \frac{1}{2}r^{2}-1-2\frac{\partial^{2}}{\partial\eta\partial
\eta^{\ast}}\right)  \left\vert \eta\right\rangle \nonumber\\
&  =\left[  \frac{r^{2}}{2}-1-\frac{1}{2}\left(  \frac{\partial^{2}}{\partial
r^{2}}+\frac{1}{r}\frac{\partial}{\partial r}+\frac{1}{^{r^{2}}}\frac
{\partial^{2}}{\partial\varphi^{2}}\right)  \right]  \left\vert \eta
\right\rangle \nonumber\\
(A_{+}^{\dagger}A_{+}-A_{-}^{\dagger}A_{-})\left\vert \eta\right\rangle  &
=\left(  \eta\frac{\partial}{\partial\eta}-\eta^{\ast}\frac{\partial}%
{\partial\eta^{\ast}}\right)  \left\vert \eta\right\rangle =-\mathtt{i}%
\frac{\partial}{\partial\varphi}\left\vert \eta\right\rangle . \label{18}%
\end{align}
Projecting Eqs.(\ref{7}) onto the $\left\langle \eta\right\vert $
representation and using (\ref{14}), (\ref{16b})-(\cite{17}), one can obtain
the following equations%
\begin{equation}
l\left\langle \eta\right\vert \left.  n,l\right\rangle =i\frac{\partial
}{\partial\varphi}\left\langle \eta\right\vert \left.  n,l\right\rangle ,
\label{21}%
\end{equation}
and%
\begin{equation}
n\left\langle \eta\right\vert \left.  n,l\right\rangle =\left[  \frac{r^{2}%
}{2}-1-\frac{1}{2}\left(  \frac{\partial^{2}}{\partial r^{2}}+\frac{1}{r}%
\frac{\partial}{\partial r}+\frac{1}{^{r^{2}}}\frac{\partial^{2}}%
{\partial\varphi^{2}}\right)  \right]  \left\langle \eta\right\vert \left.
n,l\right\rangle . \label{20}%
\end{equation}
Eq.(\ref{21}) indicates that $\left\langle \eta\right\vert \left.
n,l\right\rangle \propto e^{-il\varphi}.$\ From the uniqueness of wave
function, $e^{-il\varphi}|_{\varphi=0}=e^{-il\varphi}|_{\varphi=2\pi},$ we
know $l=0,$ $\pm1,$ $\pm2\cdots.$ So letting $\left\langle \eta\right\vert
\left.  n,l\right\rangle =R(r)e^{-il\varphi}$ and substituting it into
(\ref{20}) yields
\begin{equation}
\frac{d^{2}R}{dr^{2}}+\frac{1}{r}\frac{dR}{dr}+\left(  -r^{2}+2\left(
n+1\right)  -\frac{l^{2}}{^{r^{2}}}\right)  R=0. \label{22}%
\end{equation}
Introducing $\xi=r^{2}$ such that
\begin{equation}
\frac{dR}{dr}=2\sqrt{\xi}\frac{dR}{d\xi},\text{ }\frac{d^{2}R}{dr^{2}}%
=2\frac{dR}{d\xi}+4\xi\frac{d^{2}R}{d\xi^{2}}, \label{23}%
\end{equation}
Eq. (\ref{22}) becomes%
\begin{equation}
\frac{d^{2}R}{d\xi^{2}}+\frac{1}{\xi}\frac{dR}{d\xi}+\left(  -\frac{1}%
{4}+\frac{n+1}{2\xi}-\frac{l^{2}}{4\xi^{2}}\right)  R=0. \label{24}%
\end{equation}
Then make the variable transform in (\ref{24})
\begin{equation}
R(\xi)=e^{-\xi/2}\xi^{\left\vert l\right\vert /2}u(\xi), \label{26}%
\end{equation}
one can obtain the equation for $u(\xi),$
\begin{equation}
\xi\frac{d^{2}u}{d\xi^{2}}+(\left\vert l\right\vert +1-\xi)\frac{du}{d\xi
}+\frac{n-\left\vert l\right\vert }{2}u=0. \label{27}%
\end{equation}
Eq.(\ref{27}) is just a confluent hypergeometric equation whose solution is
the associate Laguerre polynomials, $L_{n_{\rho}}^{\left\vert l\right\vert
}(\xi)$, where $n_{\rho}=\frac{n-\left\vert l\right\vert }{2},$ $(n_{\rho}=0,$
$1,$ $2,\cdots)$\cite{13}$.$ Thus the wave function of $\left\vert
n,l\right\rangle $ in $\left\langle \eta\right\vert $ representation is given
by
\begin{equation}
\left\langle \eta\right\vert \left.  n,l\right\rangle =C_{1}e^{-il\varphi
}e^{-\frac{1}{2}r^{2}}r^{\left\vert l\right\vert }L_{n_{\rho}}^{\left\vert
l\right\vert }(r^{2}), \label{29}%
\end{equation}
where $C_{1}$ is an integral constant. The right-hand side of Eq.(\ref{29}) is
just the LG mode, so we reach the conclusion that the wave function of
$\left\vert n,l\right\rangle $ in the entangled state representation is just
the LG mode, i.e., the LG mode gets its new physical meaning in quantum optics.

Next, we further derive the explicit expression of $\left\vert
n,l\right\rangle .$ Using the completeness relation of $\left\langle
\eta\right\vert $ (\ref{16}) and (\ref{29}), we have
\begin{align}
\left\vert n,l\right\rangle  &  =\int\frac{d^{2}\eta}{\pi}\left\vert
\eta\right\rangle \left\langle \eta\right\vert n,l\rangle\nonumber\\
&  =C_{1}\int\frac{d^{2}\eta}{\pi}e^{-\frac{1}{2}r^{2}}\left\vert
\eta\right\rangle e^{-il\varphi}r^{\left\vert l\right\vert }L_{n_{\rho}%
}^{\left\vert l\right\vert }(r^{2}). \label{30}%
\end{align}
Then noticing the relation between two-variable Hermite polynomial
\cite{14,15} and Laguerre polynomial,%
\begin{equation}
H_{m,n}\left(  \eta,\eta^{\ast}\right)  =m!\left(  -1\right)  ^{m}\eta^{\ast
}{}^{n-m}L_{m}^{n-m}\left(  \eta\eta^{\ast}\right)  , \label{31}%
\end{equation}
where $m<n,\ $and the generating function of $H_{m,n}\left(  \eta,\eta^{\ast
}\right)  $ is
\begin{equation}
H_{m,n}\left(  x,y\right)  =\left.  \frac{\partial^{m+n}}{\partial
t^{m}\partial t^{\prime n}}\exp\left[  -tt^{\prime}+tx+t^{\prime}y\right]
\right\vert _{t=t^{\prime}=0}, \label{32}%
\end{equation}
as well as using the integral formula \cite{16}%
\begin{equation}
\int\frac{d^{2}z}{\pi}\exp\left(  \zeta\left\vert z\right\vert ^{2}+\xi z+\eta
z^{\ast}\right)  =-\frac{1}{\zeta}e^{-\frac{\xi\eta}{\zeta}},\text{ Re}\left(
\xi\right)  <0, \label{33}%
\end{equation}
we can reform Eq.(\ref{30}) as (without loss of the generality, setting $l>0$
and $m_{\rho}=[n+\left\vert l\right\vert ]/2$)%
\begin{align}
\left\vert n,l\right\rangle  &  =\frac{\left(  -1\right)  ^{n_{\rho}}C_{1}%
}{n_{\rho}!}\int\frac{d^{2}\eta}{\pi}H_{n_{\rho},m_{\rho}}\left(  \eta
,\eta^{\ast}\right)  e^{-\frac{1}{2}\left\vert \eta\right\vert ^{2}}\left\vert
\eta\right\rangle \nonumber\\
&  =\frac{\left(  -1\right)  ^{n_{\rho}}C_{1}}{n_{\rho}!}\frac{\partial
^{n_{\rho}+m_{\rho}}}{\partial t^{n_{\rho}}\partial t^{\prime m_{\rho}}}%
\exp\left[  -tt^{\prime}+A_{+}^{\dagger}A_{-}^{\dagger}\right] \nonumber\\
&  \times\int\frac{d^{2}\eta}{\pi}\exp\left[  -\left\vert \eta\right\vert
^{2}+\left(  A_{+}^{\dagger}+t\right)  \eta+\left(  t^{\prime}-A_{-}^{\dagger
}\right)  \eta^{\ast}\right]  _{t=t^{\prime}=0}\left\vert 00\right\rangle
\nonumber\\
&  =\frac{\left(  -1\right)  ^{n_{\rho}}C_{1}}{n_{\rho}!}\frac{\partial
^{n_{\rho}+m_{\rho}}}{\partial t^{n_{\rho}}\partial t^{\prime m_{\rho}}}%
\exp\left[  A_{+}^{\dagger}t^{\prime}-tA_{-}^{\dagger}\right]  _{t=t^{\prime
}=0}\left\vert 00\right\rangle \nonumber\\
&  =\frac{C_{1}}{n_{\rho}!}\left(  A_{+}^{\dagger}\right)  ^{m_{\rho}}\left(
A_{-}^{\dagger}\right)  ^{n_{\rho}}\left\vert 00\right\rangle . \label{34}%
\end{align}

\section{Generation of $\left\vert n,l\right\rangle $ by Beam Splitter}

Note Eq.(\ref{12}) and
\begin{align}
A_{+}^{\dag}  &  =e^{i\frac{\pi}{2}J_{x}}a_{1}^{\dag}e^{-i\frac{\pi}{2}J_{x}%
},\text{ \ }A_{-}^{\dag}=e^{i\frac{\pi}{2}J_{x}}a_{2}^{\dag}e^{-i\frac{\pi}%
{2}J_{x}},\text{ }\nonumber\\
J_{x}  &  =\frac{1}{2}\left(  a_{1}^{\dag}a_{2}+a_{2}^{\dag}a_{1}\right)
,\text{ }\left\vert 00\right\rangle =e^{i\frac{\pi}{2}J_{x}}\left\vert
00\right\rangle , \label{35}%
\end{align}
thus Eq.(\ref{34}) can be further put into the following form%
\begin{align}
\left\vert n,l\right\rangle  &  =\frac{C_{1}}{n_{\rho}!}e^{i\frac{\pi}{2}%
J_{x}}\left(  a_{1}^{\dag}\right)  ^{m_{\rho}}\left(  a_{2}^{\dag}\right)
^{n_{\rho}}\left\vert 00\right\rangle \nonumber\\
&  =C_{1}\sqrt{\frac{m_{\rho}!}{n_{\rho}!}}e^{i\frac{\pi}{2}J_{x}}\left\vert
m_{\rho},n_{\rho}\right\rangle . \label{36}%
\end{align}
It is easy to see that the normalized constant can be chosen as $C_{1}%
=\sqrt{n_{\rho}!/m_{\rho}!},$ which further leads to
\begin{equation}
\left\vert n,l\right\rangle =e^{i\frac{\pi}{2}J_{x}}\left\vert m_{\rho
},n_{\rho}\right\rangle , \label{e37}%
\end{equation}
where $J_{x}$ can be expressed by angular momentum operators $J_{+}%
=a_{1}^{\dag}a_{2}$ and $J_{-}=a_{1}a_{2}^{\dag}$, $J_{x}=\frac{1}{2}\left(
J_{+}+J_{-}\right)  .$ $J_{+},$ $J_{-}$ and $J_{z}=\frac{1}{2}\left(
a_{1}^{\dag}a_{1}-a_{2}^{\dag}a_{2}\right)  $ make up a close SU(2) Lie algebra.

On the other hand, the beam splitter is one of the few experimentally
accessible devices that may act as an entangler. In fact, the role of a beam
splitter operator \cite{j1,j2} is expressed by
\begin{equation}
B\left(  \theta,\phi\right)  =\exp\left[  \frac{\theta}{2}\left(  a_{1}^{\dag
}a_{2}e^{i\phi}-a_{1}a_{2}^{\dag}e^{-i\phi}\right)  \right]  , \label{e38}%
\end{equation}
with the amplitude reflection and transmission coefficients $T=\cos
\frac{\theta}{2},$ $R=\sin\frac{\theta}{2}.$ The beam splitter gives the phase
difference $\phi$ between the reflected and transmitted fields. Comparing
Eq.(\ref{e38}) with $e^{i\frac{\pi}{2}J_{x}}$ leads us to choose $\theta
=\pi/2$ (corresponding to 50:50 beam splitter) and $\phi=\pi/2,$ thus
$B\left(  \pi/2,\pi/2\right)  $ is just equivalent to $e^{i\frac{\pi}{2}J_{x}%
}$ in form. This indicates that $\left\vert n,l\right\rangle $ can be
generated by acting a symmetric beam splitter with $\phi=\pi/2$ on two
independent input Fock states $\left\vert m_{\rho},n_{\rho}\right\rangle
=\left\vert m_{\rho}\right\rangle _{1}\left\vert n_{\rho}\right\rangle _{2}.$
In addition, note that $n=m_{\rho}+n_{\rho},$ i.e., when the total number of
input photons is $n,$ so the output state becomes an ($n+1$)-dimensional
entangled state \cite{j3}.

\section{The Wigner representation}

As is well-known, the Wigner quasidistribution provides with a definite phase
space distribution of quantum states and is very useful in quantum statistics
and quantum optics. In this section, we evaluate the Wigner representation of
$\left\vert n,l\right\rangle .$ According Ref.\cite{17}, the Wigner
representation of $\left\vert n,l\right\rangle $ is given by
\begin{align}
W_{\left\vert n,l\right\rangle }  &  =\left\langle n,l\right\vert \Delta
_{1}\left(  x_{1},p_{1}\right)  \Delta_{2}\left(  x_{2},p_{2}\right)
\left\vert n,l\right\rangle \nonumber\\
&  =\left\langle m_{\rho},n_{\rho}\right\vert e^{-i\frac{\pi}{2}J_{x}}%
\Delta_{1}\left(  x_{1},p_{1}\right) \nonumber\\
&  \times\Delta_{2}\left(  x_{2},p_{2}\right)  e^{i\frac{\pi}{2}J_{x}%
}\left\vert m_{\rho},n_{\rho}\right\rangle , \label{37}%
\end{align}
where $\Delta_{1}\left(  x_{1},p_{1}\right)  $ is the single-mode Wigner
operator, whose Weyl ordering form \cite{18} is
\begin{equation}
\Delta_{1}\left(  x_{1},p_{1}\right)  =%
%TCIMACRO{\QATOP{:}{:}}%
%BeginExpansion
\genfrac{}{}{0pt}{}{:}{:}%
%EndExpansion
\delta\left(  p_{1}-P_{1}\right)  \delta\left(  x_{1}-X_{1}\right)
%TCIMACRO{\QATOP{:}{:}}%
%BeginExpansion
\genfrac{}{}{0pt}{}{:}{:}%
%EndExpansion
, \label{38}%
\end{equation}
where the symbol$%
%TCIMACRO{\QATOP{:}{:}}%
%BeginExpansion
\genfrac{}{}{0pt}{}{:}{:}%
%EndExpansion%
%TCIMACRO{\QATOP{:}{:}}%
%BeginExpansion
\genfrac{}{}{0pt}{}{:}{:}%
%EndExpansion
$denotes Weyl ordering \cite{19}. Note that the order of Bose operators $a$
and $a^{\dag}$ within $%
%TCIMACRO{\QATOP{:}{:}}%
%BeginExpansion
\genfrac{}{}{0pt}{}{:}{:}%
%EndExpansion%
%TCIMACRO{\QATOP{:}{:}}%
%BeginExpansion
\genfrac{}{}{0pt}{}{:}{:}%
%EndExpansion
$ can be permitted. That is to say, even though $[a$,$a^{\dag}]$ $=1$, we can
have $%
%TCIMACRO{\QATOP{:}{:}}%
%BeginExpansion
\genfrac{}{}{0pt}{}{:}{:}%
%EndExpansion
aa^{\dag}%
%TCIMACRO{\QATOP{:}{:}}%
%BeginExpansion
\genfrac{}{}{0pt}{}{:}{:}%
%EndExpansion
=%
%TCIMACRO{\QATOP{:}{:}}%
%BeginExpansion
\genfrac{}{}{0pt}{}{:}{:}%
%EndExpansion
a^{\dag}a%
%TCIMACRO{\QATOP{:}{:}}%
%BeginExpansion
\genfrac{}{}{0pt}{}{:}{:}%
%EndExpansion
$. According to the covariance property of Weyl ordering under similar
transformations \cite{18} and
\begin{align}
e^{-i\frac{\pi}{2}J_{x}}X_{1}e^{i\frac{\pi}{2}J_{x}}  &  =\frac{1}{\sqrt{2}%
}\left(  X_{1}-P_{2}\right)  ,\text{ }\nonumber\\
e^{-i\frac{\pi}{2}J_{x}}P_{1}e^{i\frac{\pi}{2}J_{x}}  &  =\frac{1}{\sqrt{2}%
}\left(  P_{1}+X_{2}\right)  ,\nonumber\\
e^{-i\frac{\pi}{2}J_{x}}X_{2}e^{i\frac{\pi}{2}J_{x}}  &  =\frac{1}{\sqrt{2}%
}\left(  X_{2}-P_{1}\right)  ,\text{ }\nonumber\\
e^{-i\frac{\pi}{2}J_{x}}P_{2}e^{i\frac{\pi}{2}J_{x}}  &  =\frac{1}{\sqrt{2}%
}\left(  P_{2}+X_{1}\right)  , \label{39}%
\end{align}
we have
\begin{align}
&  e^{-i\frac{\pi}{2}J_{x}}\Delta_{1}\left(  x_{1},p_{1}\right)  \Delta
_{2}\left(  x_{2},p_{2}\right)  e^{i\frac{\pi}{2}J_{x}}\nonumber\\
&  =%
%TCIMACRO{\QATOP{:}{:}}%
%BeginExpansion
\genfrac{}{}{0pt}{}{:}{:}%
%EndExpansion
\delta\left(  p_{1}-\frac{P_{1}+X_{2}}{\sqrt{2}}\right)  \delta\left(
x_{1}-\frac{X_{1}-P_{2}}{\sqrt{2}}\right) \nonumber\\
&  \times\delta\left(  p_{2}-\frac{P_{2}+X_{1}}{\sqrt{2}}\right)
\delta\left(  x_{2}-\frac{X_{2}-P_{1}}{\sqrt{2}}\right)
%TCIMACRO{\QATOP{:}{:}}%
%BeginExpansion
\genfrac{}{}{0pt}{}{:}{:}%
%EndExpansion
\nonumber\\
&  =\Delta_{1}\left(  \frac{x_{1}+p_{2}}{\sqrt{2}},\frac{p_{1}-x_{2}}{\sqrt
{2}}\right)  \Delta_{2}\left(  \frac{x_{2}+p_{1}}{\sqrt{2}},\frac{p_{2}-x_{1}%
}{\sqrt{2}}\right)  . \label{40}%
\end{align}
Since the Wigner representation of number state $\left\vert m\right\rangle $
is well known \cite{20},
\begin{align}
W_{\left\vert m\right\rangle }  &  =\left\langle m\right\vert \Delta
_{1}\left(  x_{1},p_{1}\right)  \left\vert m\right\rangle \nonumber\\
&  =\frac{\left(  -1\right)  ^{m}}{\pi}e^{-\left(  x_{1}^{2}+p_{1}^{2}\right)
}L_{m}\left[  2\left(  x_{1}^{2}+p_{1}^{2}\right)  \right]  , \label{41}%
\end{align}
so we directly obtain the Wigner representation of L-G mode,
\begin{align}
W_{\left\vert n,l\right\rangle }  &  =\left\langle m_{\rho}\right\vert
\Delta_{1}\left(  \frac{x_{1}+p_{2}}{\sqrt{2}},\frac{p_{1}-x_{2}}{\sqrt{2}%
}\right)  \left\vert m_{\rho}\right\rangle \nonumber\\
&  \times\left\langle n_{\rho}\right\vert \Delta_{2}\left(  \frac{x_{2}+p_{1}%
}{\sqrt{2}},\frac{p_{2}-x_{1}}{\sqrt{2}}\right)  \left\vert n_{\rho
}\right\rangle \nonumber\\
&  =\frac{\left(  -1\right)  ^{m_{\rho}+n_{\rho}}}{\pi^{2}}e^{-Q_{0}%
}L_{m_{\rho}}\left(  Q_{0}+Q_{2}\right)  L_{n_{\rho}}\left(  Q_{0}%
-Q_{2}\right)  , \label{42}%
\end{align}
where $Q_{0}=\allowbreak p_{1}^{2}+p_{2}^{2}+x_{1}^{2}+x_{2}^{2}$ and
$Q_{2}=2p_{2}x_{1}-2p_{1}x_{2}$. Eq.(\ref{42}) is in agreement with the result
of Ref. \cite{5,7}. Our derivation seems economical.

\section{The marginal distributions and fractional Fourier transform of
$\left\vert n,l\right\rangle $}

The fractional Fourier transform (FrFT) has been paid more and more attention
within different contexts of both mathematics and physics. It is also very
useful tool in Fourier optics and information optics. In this section, we
examine the relation between the FrFT and the marginal distributions of
$W_{\left\vert n,l\right\rangle }$.

For this purpose, we recall that the two-mode Wigner operator $\Delta
_{1}\left(  x_{1},p_{1}\right)  \Delta_{2}\left(  x_{2},p_{2}\right)
\equiv\Delta_{1}\left(  \alpha\right)  \Delta_{2}\left(  \beta\right)  $
($\alpha=(x_{1}+ip_{1})/\sqrt{2}$, $\beta=(x_{1}+ip_{1})/\sqrt{2}$) in
entangled state representation $\left\langle \tau\right\vert .$ Using the IWOP
technique we have shown in \cite{19} that $\Delta_{1,2}\left(  \sigma
,\gamma\right)  $ is just the product of two independent single-mode Wigner
operators $\Delta_{1}\left(  \alpha\right)  \Delta_{2}\left(  \beta\right)
=\Delta_{1,2}\left(  \sigma,\gamma\right)  $ i.e.,
\begin{equation}
\Delta_{1,2}\left(  \sigma,\gamma\right)  =\int\frac{d^{2}\tau}{\pi^{3}%
}\left\vert \sigma-\tau\right\rangle \left\langle \sigma+\tau\right\vert
e^{\tau\gamma^{\ast}-\tau^{\ast}\gamma},\label{43}%
\end{equation}
where $\sigma=\alpha-\beta^{\ast},$ $\gamma=\alpha+\beta^{\ast}$ and
$\left\vert \tau=\tau_{1}+i\tau_{2}\right\rangle $ can be expressed in
two-mode Fock space as \cite{12,13}
\begin{equation}
\left\vert \tau\right\rangle =\exp\left\{  -\frac{1}{2}\left\vert
\tau\right\vert ^{2}+\tau a_{1}^{\dagger}-\tau^{\ast}a_{2}^{\dagger}%
+a_{1}^{\dagger}a_{2}^{\dagger}\right\}  \left\vert 00\right\rangle
,\label{44}%
\end{equation}
which is the common eigenvector of $X_{1}-X_{2}$\ \ and \ $P_{1}+P_{2}$, which
obeys the eigenvector equations $\left(  X_{1}-X_{2}\right)  \left\vert
\tau\right\rangle =\sqrt{2}\tau_{1}\left\vert \tau\right\rangle ,$\ $\left(
P_{1}+P_{2}\right)  \left\vert \tau\right\rangle =\sqrt{2}\tau_{2}\left\vert
\tau\right\rangle .$

Performing the integration of $\Delta_{1,2}\left(  \sigma,\gamma\right)  $
over $d^{2}\gamma$ ($d^{2}\sigma$) leads to the projection operator of the
entangled state $\left\vert \tau\right\rangle $ ($\left\vert \xi\right\rangle
$)
\begin{align}
\int d^{2}\gamma\Delta_{1,2}(\sigma,\gamma)  &  =\frac{1}{\pi}\left\vert
\tau\right\rangle \left\langle \tau\right\vert |_{\tau=\sigma},\nonumber\\
\int d^{2}\sigma\Delta_{1,2}(\sigma,\gamma)  &  =\frac{1}{\pi}\left\vert
\xi\right\rangle \left\langle \xi\right\vert |_{\xi=\gamma}, \label{45}%
\end{align}
where $\left\vert \xi\right\rangle $ is the conjugate state of $\left\vert
\tau\right\rangle $. Thus the marginal distributions for quantum states $\rho$
in ($\tau_{1},\tau_{2}$) and ($\xi_{1},\xi_{2}$) phase space are given by
\begin{align}
\int d^{2}\sigma W\left(  \sigma,\gamma\right)   &  =\frac{1}{\pi}\left\langle
\xi\right\vert \rho\left\vert \xi\right\rangle |_{\xi=\gamma},\text{
}\nonumber\\
\int d^{2}\gamma W\left(  \sigma,\gamma\right)   &  =\frac{1}{\pi}\left\langle
\tau\right\vert \rho\left\vert \tau\right\rangle |_{\tau=\sigma}, \label{47}%
\end{align}
respectively. Eq.(\ref{47}) shows that, for bipartite system, the marginal
distributions can be calculated by evaluating the quantum average of $\rho$ in
$\left\langle \xi\right\vert $, $\left\langle \tau\right\vert $ representations.

Now we calculate the inner-product $\left\langle \tau\right.  \left\vert
n,l\right\rangle .$ Note that $a_{1}^{\dag}a_{2}$, $a_{1}a_{2}^{\dag}$, and
$J_{z}=\frac{1}{2}\left(  a_{1}^{\dag}a_{1}-a_{2}^{\dag}a_{2}\right)  $ make
up a close SU(2) Lie algebra, thus $e^{i\frac{\pi}{2}J_{x}}$ can be decomposed
as
\begin{equation}
e^{i\frac{\pi}{2}J_{x}}=e^{ia_{1}^{\dag}a_{2}}\exp\left[  \frac{1}{2}\left(
a_{1}^{\dag}a_{1}-a_{2}^{\dag}a_{2}\right)  \ln2\right]  e^{ia_{2}^{\dag}%
a_{1}}, \label{50}%
\end{equation}
then we have%

\begin{align}
\left\langle \tau\right.  \left\vert n,l\right\rangle  &  =\sqrt{\frac
{m_{\rho}!}{n_{\rho}!}}\sum_{k=0}^{m_{\rho}}\left(  \sqrt{2}\right)
^{m_{\rho}-n_{\rho}-2k}\nonumber\\
&  \times\frac{\left(  n_{\rho}+k\right)  !}{k!\left(  m_{\rho}-k\right)
!}\sum_{j=0}^{n_{\rho}+k}\frac{i^{k+j}}{j!}\sqrt{\frac{\left(  m_{\rho
}-k+j\right)  !}{\left(  n_{\rho}+k-j\right)  !}}\nonumber\\
&  \times\left\langle \tau\right.  \left\vert m_{\rho}-k+j,n_{\rho
}+k-j\right\rangle . \label{51}%
\end{align}

Using the generating function of $H_{m,n},$ we have%
\begin{equation}
\left\langle \tau^{\prime}\right.  \left\vert m,n\right\rangle =\frac{\left(
-1\right)  ^{n}}{\sqrt{m!n!}}H_{m,n}\left(  \tau^{\prime\ast},\tau^{\prime
}\right)  e^{-\left\vert \tau^{\prime}\right\vert ^{2}/2}. \label{52}%
\end{equation}
Substituting Eq.(\ref{52}) into Eq.(\ref{51}) leads to
\begin{align}
\left\langle \tau\right.  \left\vert n,l\right\rangle  &  =\left(  -\right)
^{n_{\rho}}2^{\left(  m_{\rho}-n_{\rho}\right)  /2}e^{-\left\vert
\tau\right\vert ^{2}/2}\nonumber\\
&  \sqrt{\frac{m_{\rho}!}{n_{\rho}!}}\sum_{k=0}^{m_{\rho}}\frac{\left(
n_{\rho}+k\right)  !}{2^{k}k!\left(  m_{\rho}-k\right)  !}\nonumber\\
&  \times\sum_{j=0}^{n_{\rho}+k}\frac{\left(  -i\right)  ^{k+j}}{j!\left(
n_{\rho}+k-j\right)  !}H_{m_{\rho}-k+j,n_{\rho}+k-j}\left(  \tau^{\ast}%
,\tau\right)  . \label{53}%
\end{align}
Thus the marginal distribution is%
\begin{align}
&  \int d^{2}\gamma W\left(  \sigma,\gamma\right) \nonumber\\
&  =\frac{e^{-\left\vert \sigma\right\vert ^{2}}}{\pi}2^{m_{\rho}-n_{\rho}%
}\frac{m_{\rho}!}{n_{\rho}!}\left\vert \sum_{k=0}^{m_{\rho}}\frac{\left(
-i\right)  ^{k}\left(  n_{\rho}+k\right)  !}{2^{k}k!\left(  m_{\rho}-k\right)
!}\right. \nonumber\\
&  \times\left.  \sum_{j=0}^{n_{\rho}+k}\left(  -i\right)  ^{j}\frac
{H_{m_{\rho}-k+j,n_{\rho}+k-j}\left(  \sigma^{\ast},\sigma\right)  }{j!\left(
n_{\rho}+k-j\right)  !}\right\vert ^{2}. \label{54}%
\end{align}
Due to the presence of sum polynomial, the marginal distribution is not
Gaussian. In a similar way, one can obtain the other marginal distribution in
$\gamma$ direction.

Before the end of this section, we mention the relation between the FrFT and
marginal distribution. In Ref.\cite{j5} we have proved that, in the context of
quantum optics, the FrFT can be described as the matrix element of fractional
operator $\exp[-i\alpha(a_{1}^{\dagger}a_{1}+a_{2}^{\dagger}a_{2})]$ between
$\left\langle \tau^{\prime}\right\vert $ and $\left\vert f\right\rangle $, i.e.,%

\[
\mathcal{F}_{\alpha}\left[  f\left(  \tau^{\prime}\right)  \right]
=\frac{e^{i(\alpha-\frac{\pi}{2})}}{2\sin\alpha}\int\frac{d^{2}\tau^{\prime}%
}{\pi}\exp\left[  \frac{i(\left\vert \tau^{\prime}\right\vert ^{2}+\left\vert
\tau\right\vert ^{2})}{2\tan\alpha}-\frac{i\left(  \tau^{\ast}\tau^{\prime
}+\tau^{\prime\ast}\tau\right)  }{2\sin\alpha}\right]  f\left(  \tau^{\prime
}\right)  =\left\langle \tau\right\vert \exp\left[  -i\alpha\left(
a_{1}^{\dagger}a_{1}+a_{2}^{\dagger}a_{2}\right)  \right]  \left\vert
f\right\rangle ,
\]
where $f\left(  \tau^{\prime}\right)  =\left\langle \tau^{\prime}\right\vert
\left.  f\right\rangle .$ When $\left\vert f\right\rangle =e^{i\frac{\pi}%
{2}J_{x}}\left\vert m_{\rho},n_{\rho}\right\rangle ,$ the corresponding FrFT
is%
\begin{align}
\mathcal{F}_{\alpha}\left[  \left\langle \tau^{\prime}\right.  \left\vert
n,l\right\rangle \right]   &  =\left\langle \tau\right\vert \exp\left[
-i\alpha\left(  a_{1}^{\dagger}a_{1}+a_{2}^{\dagger}a_{2}\right)  \right]
e^{i\frac{\pi}{2}J_{x}}\left\vert m_{\rho},n_{\rho}\right\rangle \nonumber\\
&  =e^{-i\alpha\left(  m_{\rho}+n_{\rho}\right)  }\left\langle \tau\right.
\left\vert n,l\right\rangle , \label{49}%
\end{align}
where we have used $e^{-i\alpha a_{1}^{\dagger}a_{1}}a_{1}$ $e^{i\alpha
a_{1}^{\dagger}a_{1}}=a_{1}e^{i\alpha}.$ Eq.(\ref{49}) implies that the
eigenequations of FrFT can also be $\left\langle \tau\right.  \left\vert
n,l\right\rangle $ with the eigenvalue being $e^{-i\alpha\left(  m_{\rho
}+n_{\rho}\right)  }$, which is the superposition (\ref{51}) of two-variable
Hermite polynomials. In Ref.\cite{j6}, we have proved that the two-variable
Hermite polynomials (TVHP) is just the eigenfunction of the FrFT in complex
form by using the IWOP technique and the bipartite entangled state
representations. Here, we should emphasize that for any unitary two-mode
operators $U$ obeying the relation $\exp[-i\alpha(a_{1}^{\dagger}a_{1}%
+a_{2}^{\dagger}a_{2})]U\exp[i\alpha(a_{1}^{\dagger}a_{1}+a_{2}^{\dagger}%
a_{2})]=U,$ the wave function $\left\langle \tau\right\vert U\left\vert
m_{\rho},n_{\rho}\right\rangle $ is the eigenfunction of FrFT with the
eigenvalue being $e^{-i\alpha\left(  m_{\rho}+n_{\rho}\right)  }$ \cite{2}.

Combing Eqs.(\ref{49}) and (\ref{47}), one can obtain a simple formula
connecting the FrFT and the marginal distribution of $W_{\left\vert
n,l\right\rangle },$%
\begin{equation}
\int d^{2}\gamma W\left(  \sigma,\gamma\right)  =\frac{1}{\pi}\left\vert
\mathcal{F}_{\alpha}\left[  \left\langle \tau=\sigma\right.  \left\vert
n,l\right\rangle \right]  \right\vert ^{2}, \label{55}%
\end{equation}
which is $\alpha-$independent. Thus we can also obtain the marginal
distribution by the FrFT.

\section{L-G mode as generalized Wigner transform of Hermit-Gaussian modes}

In this section we shall reveal the relation between the L-G mode and the
single variable Hermit-Gaussian(H-G) modes. Note that by taking the Fourier
transformation of $\left\vert \tau\right\rangle $ with regard to $\tau_{2}$
followed by the inverse Fourier transformation, we can recover the entangled
state $\left\vert \tau\right\rangle .$ In another word, $\left\vert \tau
=\tau_{1}+i\tau_{2}\right\rangle $ can be decomposed into
\begin{align}
\left\vert \tau\right\rangle  &  =\int_{-\infty}^{\infty}dxe^{ix\sqrt{2}%
\tau_{2}}\left\vert x+\frac{\tau_{1}}{\sqrt{2}}\right\rangle _{1}%
\otimes\left\vert x-\frac{\tau_{1}}{\sqrt{2}}\right\rangle _{2}\nonumber\\
&  =e^{-i\tau_{1}\tau_{2}}\int_{-\infty}^{\infty}dxe^{ix\sqrt{2}\tau_{2}%
}\left\vert x\right\rangle _{1}\otimes\left\vert x-\sqrt{2}\tau_{1}%
\right\rangle _{2}, \label{56}%
\end{align}
in which $\left\vert x\right\rangle _{i}$ ($i=1,2$) are the coordinate
eigenvectors. Eq.(\ref{56}) is called the Schmidt decomposition of $\left\vert
\tau\right\rangle $ and indicates $\left\vert \tau\right\rangle $ is an
entangled state \cite{21}.

Eq.(\ref{56}) leads to
\begin{align}
\left\langle m,n\right.  \left\vert \tau\right\rangle  &  =\int_{-\infty
}^{\infty}dxe^{ix\sqrt{2}\tau_{2}}\left\langle m\left\vert x+\frac{\tau_{1}%
}{\sqrt{2}}\right.  \right\rangle \left\langle n\left\vert x-\frac{\tau_{1}%
}{\sqrt{2}}\right.  \right\rangle \nonumber\\
&  =\left\langle m\right\vert \left[  \int_{-\infty}^{\infty}dxe^{ix\sqrt
{2}\tau_{2}}\left\vert x+\frac{\tau_{1}}{\sqrt{2}}\right\rangle \left\langle
x-\frac{\tau_{1}}{\sqrt{2}}\right\vert \right]  \left\vert n\right\rangle .
\label{57}%
\end{align}
It is interesting to notice that the integration [...] in Eq.(\ref{57}) is
similar to the single-mode Wigner operator,%
\begin{equation}
\Delta\left(  x,p\right)  =\frac{1}{2\pi}\int_{-\infty}^{\infty}%
due^{-iup}\left\vert x-\frac{u}{2}\right\rangle \left\langle x+\frac{u}%
{2}\right\vert , \label{58}%
\end{equation}
and the left hand side of Eq.(\ref{57}) corresponds to TVHP mode (L-G mode).
It might be expected that the L-G can be expressed in terms of the generalized
Wigner transform (GWT). Actually, after making variable replacement,
Eq.(\ref{57}) can be rewritten as%
\begin{equation}
\left\langle m,n\right.  \left\vert \tau\right\rangle =\pi\left\langle
m\right\vert \Delta\left(  \frac{\tau_{1}}{\sqrt{2}},\frac{\tau_{2}}{\sqrt{2}%
}\right)  \left(  -1\right)  ^{N}\left\vert n\right\rangle . \label{59}%
\end{equation}
If we introduce the following GWT,
\begin{equation}
W_{g}\left[  f,v\right]  \left(  x,p\right)  =\left\langle f\right\vert
\Delta\left(  x,p\right)  \left\vert v\right\rangle , \label{60}%
\end{equation}
which reduces to the usual Wigner transform under the condition $\left\vert
v\right\rangle =\left\vert f\right\rangle $, while for $\left\vert
v\right\rangle =\left(  -1\right)  ^{n}\left\vert n\right\rangle $ and
$\left\vert f\right\rangle =\left\vert m\right\rangle $ Eq.(\ref{60}) becomes
the right hand side of Eq.(\ref{59}), which corresponds to the GWT.

On the other hand, note that Eqs. (\ref{31}) and (\ref{52}), the left hand
side of Eq.(\ref{59}) can be put into (without loss of the generality, letting
$m<n$)
\begin{align}
\left\langle m,n\right.  \left\vert \tau\right\rangle  &  =\frac{\left(
-1\right)  ^{n}}{\sqrt{m!n!}}H_{m,n}\left(  \tau,\tau^{\ast}\right)
e^{-\left\vert \tau\right\vert ^{2}/2}\nonumber\\
&  =\left(  -1\right)  ^{n+m}\sqrt{\frac{m!}{n!}}\tau^{\ast}{}^{n-m}%
L_{m}^{n-m}\left(  \tau\tau^{\ast}\right)  e^{-\left\vert \tau\right\vert
^{2}/2}, \label{61}%
\end{align}
which indicates that the left hand side of Eq.(\ref{59}) is just corresponding
to the L-G mode, as well as%
\begin{equation}
\left\langle m\right.  \left\vert x\right\rangle =\frac{e^{-x^{2}/2}}%
{\sqrt{2^{m}m!\sqrt{\pi}}}H_{m}\left(  x\right)  \equiv h_{m}\left(  x\right)
, \label{62}%
\end{equation}
where $H_{m}\left(  x\right)  $ is single variable Hermite polynomial, and
$h_{m}\left(  x\right)  $ just corresponds to the H-G mode, we have
\begin{align}
&  \sqrt{\frac{m!}{n!}}\tau^{\ast}{}^{n-m}L_{m}^{n-m}\left(  \tau\tau^{\ast
}\right)  e^{-\left\vert \tau\right\vert ^{2}/2}\nonumber\\
&  =\frac{\left(  -1\right)  ^{m}}{2}\int_{-\infty}^{\infty}due^{-iu\frac
{\tau_{2}}{\sqrt{2}}}h_{m}\left(  \frac{\tau_{1}}{\sqrt{2}}-\frac{u}%
{2}\right)  h_{n}\left(  \frac{\tau_{1}}{\sqrt{2}}+\frac{u}{2}\right)
\nonumber\\
&  =\left(  -1\right)  ^{m}\pi W_{g}\left[  h_{m},h_{n}\right]  \left(
\tau_{1}/\sqrt{2},\tau_{2}/\sqrt{2}\right)  . \label{63}%
\end{align}
Thus, we can conclude that the L-G mode can be obtained by the GWT of two
single-variable H-G modes. In addition, we should point out that the L-G mode
can also be generated by windowed Fourier transform (which is often used in
signal process) of two single-variable H-G modes by noticing the second line
of Eq.(\ref{56}).

In summary, we have endowed the Laguerre-Gaussian (LG) mode with new physical
meaning in quantum optics, i.e., we find that it is just the wave function of
the common eigenvector of the orbital angular momentum and the total photon
number operators of 2-d oscillator in the entangled state representation. The
common eigenvector can be obtained by using beam splitter with the phase
difference $\phi=\pi/2$ between the reflected and transmitted fields. With the
aid of the Weyl ordering invariance under similar transforms, the Wigner
representation of LG is directly obtained. It is shown that its marginal
distributions can be calculated by the FrFT. In addition, L-G mode can also be
considered as the generalized Wigner transform of Hermite Gaussian modes by
using the Schmidt decomposition of the entangled state representation.

\textbf{ACKNOWLEDGEMENT: }Work supported by a grant from the Key Programs
Foundation of Ministry of Education of China (No. 210115) and the Research
Foundation of the Education Department of Jiangxi Province of China (No.
GJJ10097). L.-Y. Hu's email address is hlyun2008@126.com.

\end{document}